\documentstyle{aipproc2}
\begin{document}
\title{GRAVITATIONAL MICROLENSING AND DARK MATTER IN THE GALACTIC HALO}

\author{PHILIPPE JETZER}
\address{Paul Scherrer Institute, Laboratory for Astrophysics, CH-5232 
Villigen PSI, and
Institute of Theoretical Physics, University of Z\"urich, 
Winterthurerstrasse
190, CH-8057 Z\"urich, Switzerland}

\maketitle

\begin{abstract}
We present the basics of microlensing and give an overview of the 
results obtained so far.
We also describe a scenario in which dark clusters of 
MACHOs (Massive Astrophysical Compact Halo Objects) and 
cold molecular clouds (mainly of $H_2$) naturally form in the halo at 
galactocentric distances larger than 10-20 kpc. 
Moreover, we discuss various 
experimental tests of this picture in particular a $\gamma$-ray
emission from the clouds due to the scattering of high-energy
cosmic-ray protons. Our estimate for the
$\gamma$-ray flux turns out to be
in remarkably good agreement 
with the recent discovery by Dixon et al. \cite{dixon}
of a possible $\gamma$-ray emission from the halo
using EGRET data.
\end{abstract}

\section*{Introduction}
A central problem in astrophysics concerns the nature of the
dark matter in galactic halos, whose presence is implied by the flat 
rotation curves in spiral galaxies. As first proposed by 
Paczy\'nski \cite{kn:Paczynski}, 
gravitational microlensing can provide a decisive answer to 
that question, and since 1993 this dream has started to become a reality 
with the detection of several microlensing events towards the Large 
Magellanic Cloud. Today, although the evidence for 
Massive Astrophysical Compact Halo Objects (MACHOs) is firm, the 
implications of this discovery crucially depend on the assumed galactic 
model. 
Moreover, at least two of the events found towards the Large Magellanic Clouds
are due to lenses located in the Clouds themselves. Therefore, it might well 
be that also the other events or at least a fraction of them
are due to MACHOs in the Clouds.
This issue might be solved when more events will be available.

It has become customary to take the standard spherical halo model as 
a baseline for comparison. Within this model, the mass moment method yields 
an average MACHO mass of \cite{kn:je} $0.27~M_{\odot}$. 
Unfortunately, because of the presently available limited statistics 
different data-analysis procedures
lead to results which are only marginally consistent. 
For instance, the average MACHO mass reported by the MACHO team 
based on its first two years of data is
\cite{kn:Pratt}
$0.5^{+0.3}_{-0.2}~M_{\odot}$. Apart from the low-statistics problem  --
which will automatically disappear from future larger data samples -- we 
feel that the real question is whether the 
standard spherical halo model correctly describes our galaxy.
Although the answer was believed to lie in the affirmative for some years, 
nowadays various arguments strongly favour a nonstandard galactic halo.
Indeed, besides the observational evidence that spiral galaxies generally have 
flattened halos, recent determinations of both the disk scale length, and the 
magnitude and slope of the rotation at the solar position indicate that
our galaxy is best described by the maximal disk model.
This conclusion is further strengthened 
by the microlensing results towards the galactic centre, which 
imply that the bulge is more massive than previously thought.
Correspondingly, the halo plays a less dominant role 
than within the standard halo model, thereby reducing the halo microlensing 
rate as well as the average MACHO mass. A similar result occurs within
the King-Michie halo models \cite{kn:Ingrosso}, which take
into account the finite escape velocity and the anisotropies in velocity 
space (typically arising during the phase of halo formation). Moreover, 
practically the same conclusions also hold  for flattened galactic models 
with a substantial degree of halo rotation. So, the expected average MACHO 
mass should be smaller than within the standard halo model. 
Still, the problem remains to explain the 
formation of MACHOs, as well as the nature of the remaining dark matter in 
galactic halos.

We have proposed a scenario \cite{kn:de1,kn:de2,kn:de4}
in which dark clusters of MACHOs and 
cold molecular clouds -- mainly of $H_2$ -- naturally form in the halo at 
galactocentric distances larger than $10-20$ kpc (somewhat 
similar ideas have also 
been put forward by  Carr and Ashman \cite{kn:carr,kn:ashman}, Pfenniger,
Combes and Martinet \cite{kn:Pfenniger},  
Gerhard and Silk
\cite{kn:Silk1} and by Fabian and Nulsen \cite{kn:fabian}). 

Here, we discuss the dark matter problem in the halo of our
Galaxy in connection with 
microlensing searches and we briefly review the main features of our 
scenario, along with its observational implications in particular with 
a $\gamma$-ray flux produced in the scattering of high-energy
cosmic-ray protons on $H_2$. Our estimate for the 
halo $\gamma$-ray flux turns out to be
in remarkably good agreement 
with the recent
discovery by Dixon et al. \cite{dixon} of a possible
$\gamma$-ray emission from the halo using EGRET data.
  
The content is as follows:
first, we review the evidence for dark matter 
in the halo of our Galaxy. 
As next we present the baryonic candidates for dark matter
and 
we discuss the basics of microlensing
(optical depth, microlensing rates, etc.).
We then give an overview of the results of
microlensing searches achieved so far and
we briefly present 
a scenario in which part of the dark matter
is in the form of cold molecular clouds (mainly of $H_2$).

\section*{Mass of the Milky Way} 

The best evidence for dark matter in galaxies comes from the rotation
curves of spirals.
Measurements of the rotation velocity $v_{rot}$ of stars up to the 
visible edge 
of the spiral galaxies and of $HI$ gas in the disk beyond the
optical radius (by measuring the 
Doppler shift in the 21-cm line) imply that $v_{rot} \approx$ constant
out to very large distances, rather than to show a Keplerian falloff.
These observations started around 1970 \cite{kn:Rubin}, thanks
to the improved sensitivity in both optical and 21-cm bands.
By now there are observations for over thousand 
spiral galaxies with reliable
rotation curves out to large radii. In almost all of them
the rotation curve is flat or slowly rising out to the last measured
point. Very few galaxies show falling rotation curves and those
that do either fall less rapidly than Keplerian have nearby companions
that may perturb the velocity field or have large spheroids
that may increase the rotation velocity near the centre.

There are also measurements of the rotation velocity for our Galaxy.
However, these observations turn out to be rather difficult, and
the rotation curve has been measured only up to a distance of about
20 kpc. Without any doubt our own galaxy has a typical flat 
rotation curve.
A fact this, which implies 
that it is possible to search directly for dark matter
characteristic of spiral galaxies in our own Milky Way.

In oder to infer the total mass one can also study the proper
motion of the Magellanic Clouds and of other satellites of our
Galaxy.
Recent studies \cite{kn:Zaritsky,kn:Lin,kn:Kochanek}
do not yet allow an accurate 
determination of $v_{rot}(LMC)/v_0$ 
($v_0 = 210 \pm 10$ km/s  being the local rotational velocity).
Lin et al. \cite{kn:Lin}  
analyzed the proper motion observations and concluded that 
within 100 kpc the Galactic halo has a mass 
$\sim 5.5 \pm 1 \times 10^{11} M_{\odot}$ and a substantial fraction 
$\sim 50\%$ of this mass is distributed beyond the present distance
of the Magellanic Clouds of about 50 kpc. Beyond 100 kpc the mass may 
continue to increase to $\sim 10^{12} M_{\odot}$ within its tidal radius
of about 300 kpc. This value for the total mass of the Galaxy is in
agreement with the results of Zaritsky et al. \cite{kn:Zaritsky}, who found
a total mass in the range 9.3 to 12.5 $\times 10^{11} M_{\odot}$, the former
value by assuming radial satellite orbits whereas the latter by assuming
isotropic satellite orbits.
 
The results of Lin et al. \cite{kn:Lin} suggest that
the mass of the halo dark matter up to the Large Magellanic Cloud
(LMC) is roughly half of the value
one gets for the standard halo model (with flat rotation
curve up to the LMC and spherical shape), implying thus the same reduction
for the number of expected microlensing events.
Kochanek \cite{kn:Kochanek} analysed the global mass distribution of the
Galaxy adopting a Jaffe model, whose parameters are determined
using the observations on the
proper motion of the satellites of the Galaxy, the Local
Group timing constraint and the ellipticity of the M31 orbit. 
With these observations Kochanek \cite{kn:Kochanek}
concludes that the mass inside 50 kpc is $5.4 \pm 1.3 \times 
10^{11} M_{\odot}$.
This value becomes, however, slightly smaller when using only the satellite 
observations and the disk rotation constraint, in this case
the median mass interior to 50 kpc is in the interval 3.3 to 6.1
(4.2 to 6.8) without (with) Leo I satellite in units of $10^{11} M_{\odot}$.
The lower bound without Leo I is 65\% of the mass expected assuming
a flat rotation curve up to the LMC.

\section*{Baryonic dark matter candidates}

Before discussing the baryonic dark matter
we would like to mention that another
class of candidates which is seriously taken into consideration
is the so-called cold dark matter, which
consists for instance of axions
or supersymmetric particles like neutralinos \cite{kn:jungman}.
Here, we will not discuss cold dark matter
in detail. However, recent studies
seem to point out that there is a discrepancy between the calculated (through
N-body simulations)
rotation curve for dwarf galaxies assuming an halo of cold dark matter
and the measured curves \cite{kn:moore,kn:navarro,kn:Silk}. 
If this fact is confirmed, this
would exclude cold dark matter as a major constituent of the 
halo of dwarf galaxies and possibly also of spiral
galaxies.

From the Big Bang nucleosynthesis model \cite{kn:copi,kn:PDG} 
and from the observed 
abundances of primordial elements one infers:
$0.010 \leq h^2_0 \Omega_B \leq 0.016$ or
with $h_0 \simeq 0.4 - 1$ one gets $0.01 \leq \Omega_B \leq 0.10$
(where $\Omega_B = \rho_B /\rho_{crit}$, and $\rho_{crit}=3H_0^2/8\pi G$).
Since for the amount of luminous baryons one finds
$\Omega_{lum} \ll \Omega_B$, it follows that an
important fraction 
of the baryons are dark.
Indeed, the dark baryons may well make up the entire dark halo matter.

The halo dark matter cannot be in the form of hot 
ionized hydrogen gas otherwise there would be a large
$X$-ray flux, for which there are stringent upper limits \cite{kn:corx}.
The abundance of neutral hydrogen gas
is inferred from the 21-cm measurements, which show that its contribution is 
small. Another possibility is that the hydrogen gas is in molecular form
clumped into cold clouds, as  
we will discuss later on.
Baryons could otherwise have been processed in stellar remnants
(for a detailed discussion see \cite{kn:Carr}).
If their mass is below $\sim0.08~M_{\odot}$ they are too light to ignite
hydrogen burning reactions. 
The possible origin of such brown dwarfs or Jupiter like bodies
(called also MACHOs),
by fragmentation or by some other mechanism, is at present
not well understood.  It has also been pointed out that the mass distribution
of the MACHOs, normalized to the dark halo mass density, could be 
a smooth continuation of the known initial mass function 
of ordinary stars
\cite{kn:Derujula1}. 
The ambient radiation, or their own body heat, would make
sufficiently small objects of H and He evaporate rapidly.
The condition that the rate of evaporation of such a hydrogenoid sphere be
insufficient to halve its mass in a billion years leads to the 
following lower limit on their mass \cite{kn:Derujula1}: 
$M > 10^{-7} M_{\odot}(T_S /30~ K)^{3/2} (1~ g~cm^{-3}/ \rho)^{1/2}$
($T_S$ being their surface
temperature and $\rho$ their average density, which we expect
to be of the order $\sim 1~ g~ cm^{-3}$).

Otherwise, 
MACHOs might be M-dwarfs or white dwarfs.
As a matter of fact, a deeper analysis shows that the M-dwarf option
looks problematic. The null result of several searches for low-mass stars
both in the disk and in the halo of our
Galaxy suggests that the halo cannot be mostly in the form of hydrogen
burning main sequence M-dwarfs. Optical imaging of high-latitude
fields taken with the Wide Field Camera of the Hubble Space Telescope
indicates that less than $\sim 6\%$ of the halo can be in this
form \cite{kn:JBahcall}. 
However, these results are derived under the assumption
of a smooth spatial distribution of M-dwarfs, and become considerably
less severe in the case of a clumpy distribution \cite{kn:Kerins1}.

A scenario
with white dwarfs as a major constituent of the galactic halo
dark matter has been explored \cite{kn:Tamanaha}.
However, it requires a rather ad hoc initial mass function sharply 
peaked around 2 - 6 $M_{\odot}$. Future
Hubble deep field exposures could either find the white dwarfs 
or put constraints on their fraction in the halo \cite{kn:Kawaler}.
Also a substantial component of neutron
stars and black holes with mass higher than $\sim 1~M_{\odot}$ 
is excluded, for otherwise they would  lead to an overproduction of heavy 
elements relative to the observed abundances.

\section*{Basics of microlensing}

In the following we present the main features of microlensing,
in particular its probability and rate of events 
(for reviews see also \cite{kn:Pac,kn:Roulet,kn:napoli}, whereas for double
lenses see for instance ref. \cite{kn:Dominik}).
An important issue is 
the determination from the observations of the mass of the MACHOs that
acted as gra\-vi\-tational lenses as well as the fraction of halo dark
matter they make up.
The most appropriate way to compute the average mass and other
important information is to use
the method of mass moments developed by De R\'ujula et al. \cite{kn:Derujula}.

\subsection*{Microlensing probability}

When a
MACHO of mass $M$ is sufficiently close to the line of sight
between us and a more distant
star, the light from the source suffers a gravitational
deflection. 
The deflection angle is usually so small that we do not see
two images but rather a magnification  of the original star brightness.
This magnification, at its maximum, is given by
\begin{equation}
A_{max}=\frac{u^2+2}{u(u^2+4)^{1/2}}~ . \label{eq:bb}
\end{equation}
Here $u=d/R_E$ ($d$ is the distance of the MACHO from the line of sight)
and the Einstein radius $R_E$ is defined as
\begin{equation}
R_E^2=\frac{4GMD}{c^2}x(1-x) \label{eq:cc}
\end{equation}
with $x=s/D$, and
where $D$ and $s$ are the distance between the source, respectively 
the MACHO and the observer. 

An important quantity is the optical depth $\tau_{opt}$ 
to gravitational microlensing defined as
\begin{equation}
\tau_{opt}=\int_0^1 dx \frac{4\pi G}{c^2}\rho(x) D^2 x(1-x)
\label{eq:za}
\end{equation}
with $\rho(x)$ the mass density of microlensing matter at distance
$s=xD$ from us along the line of sight. 
The quantity $\tau_{opt}$ is the probability
that a source is found within a radius $R_E$ of some MACHO and thus has a
magnification that is larger
than $A= 1.34$ ($d \leq R_E$).

We calculate $\tau_{opt}$ for a galactic mass
distribution of the form
\begin{equation}
\rho(\vec r)=\frac{\rho_0(a^2+R^2_{GC})}
{a^2+\vec r^2}~, \label{eq:zb}
\end{equation}
$\mid \vec r \mid$ being the distance from the Earth.
Here, $a$ is the core radius,
$\rho_0$ the local dark mass
density in the solar system and $R_{GC}$ the distance
between the observer and the Galactic centre.
Standard values for the
parameters are
$\rho_0=0.3~GeV/cm^3=7.9~10^{-3} M_\odot/pc^3$,
$a=5.6~kpc$ and $R_{GC}=8.5~kpc$.
With these values we get, for a spherical halo, $\tau_{opt} \simeq
5 \times 10^{-7}$
for the LMC and $\tau_{opt} \simeq 7\times 10^{-7}$ 
for the SMC \cite{kn:locarno}.

The magnification of the brightness of a star by a MACHO is a time-dependent
effect.
For a source that can be considered as
pointlike (this is the case if the projected star radius at the MACHO
distance is much less than $R_E$) 
the light curve as a function of time is obtained by inserting
\begin{equation}
u(t)=\frac{(d^2+v^2_Tt^2)^{1/2}}{R_E} \label{eq:zd}
\end{equation}
into eq.(\ref{eq:bb}), 
where $v_T$ is the transverse velocity of the MACHO, which can be inferred
from the measured rotation curve ($v_T \approx 200~ km/s$). The
achromaticity, symmetry and uniqueness of the signal are distinctive
features that allow to discriminate a microlensing event from
background events such as variable stars.

The behaviour of the magnification with time, $A(t)$, determines two
observables namely, the magnification at the peak $A(0)$ - denoted
by $A_{max}$ -
and the width of the signal $T$ (defined as 
being $T = R_E/v_T$).

\subsection*{Microlensing rate towards the LMC}

The microlensing rate depends on the mass and velocity distribution of
MACHOs. 
The mass density at a distance $s=xD$ from the observer is given by
eq.(\ref{eq:zb}).
The isothermal
spherical halo model does not determine the MACHO number density as a
function of mass. A
simplifying  assumption is to let the mass distribution be independent
of the position in the galactic halo, i.e., we assume the following
factorized form for the number density per unit mass $dn/dM$,
\begin{equation}
\frac{dn}{dM}dM=\frac{dn_0}{d\mu}
\frac{a^2+R_{GC}^2}{a^2+R_{GC}^2+D^2x^2-2DR_{GC}x cos\alpha}~d\mu=
\frac{dn_0}{d\mu} H(x) d\mu~,
\label{eq:zj}
\end{equation}
with $\mu=M/M_{\odot}$ ($\alpha$ is the angle of the 
line of sight with 
the direction of the galactic centre, which is $82^0$ for
the LMC), $n_0$ not depending on $x$ 
and is subject to the normalization
$\int d\mu \frac{dn_0}{d\mu}M=\rho_0$.
Nothing a priori is known on the distribution $d n_0/dM$.

A different situation arises for the velocity
distribution in the isothermal
spherical halo model, its
projection in the plane perpendicular to the line of sight
leads to the following
distribution in the transverse velocity $v_T$
\begin{equation}
f(v_T)=\frac{2}{v_H^2}v_T e^{-v^2_T/v_H^2} \label{eq:zr}
\end{equation}
($v_H \approx 210~km/s$ is the observed velocity dispersion in the halo).

In order to find the rate at which a single star
is microlensed with magnification
$A \geq A_{min}$, we consider MACHOs
with masses between $\mu$ and $\mu+\delta \mu$, located at a distance from
the observer between $x$ and $x+\delta x$ and with transverse velocity
between $v_T$ and $v_T+\delta v_T$. The collision time can be
calculated using the well-known fact that the inverse of the collision
time is the product of the MACHO number density, the microlensing
cross-section and the velocity. 
The rate $d\Gamma$, taken also as a differential with respect 
to the variable $u$, at which a single star is microlensed
in the interval $d\mu du dv_T dx$ is given by
\cite{kn:Derujula,kn:Griest1}
\begin{equation}
d\Gamma=2v_T f(v_T)D r_E [\mu x(1-x)]^{1/2} H(x)
\frac{d n_0}{d\mu}d\mu du dv_T dx,\label{eq:zt}
\end{equation}
with
\begin{equation}
r_E^2=\frac{4GM_{\odot}D}{c^2} \sim
(3.2\times 10^9 km)^2 .\label{eq:zs}
\end{equation}

One has to integrate
the differential number of microlensing events, 
$dN_{ev}=N_{\star} t_{obs} d\Gamma$,
over an appropriate range for $\mu$, $x$,
$u$ and $v_T$, 
in order to obtain the total number of microlensing events which can
be compared with an experiment
monitoring $N_{\star}$ stars during an
observation time $t_{obs}$ and which is able to detect
a magnification such that $A_{max} \geq A_{TH}$.
The limits of the $u$ integration are determined by
the experimental threshold in magnitude shift, $\Delta m_{TH}$:
we have $0 \leq u \leq u_{TH}$.

The range of integration for $\mu$ is where the mass
distribution $dn_0/d\mu$ is not vanishing
and that for $x$ is
$0\leq x \leq D_h/D$ where $D_h$ is the extent of the galactic halo along
the line of sight (in the case of the LMC,
the star is inside the galactic halo and thus $D_h/D=1$.)
The galactic velocity distribution is cut at the escape velocity
$v_e \approx 640~km/s$ and therefore
$v_T$ ranges over $0\leq v_T \leq v_e$.
In order to simplify the integration we integrate $v_T$
over all the positive axis, due to the exponential factor in $f(v_T)$ the
so committed error is negligible.

However, the integration range of $d\mu du dv_T dx$
does not span all the interval we have just described.
Indeed, each experiment has time
thresholds $T_{min}$ and $T_{max}$ and only detects events with:
$T_{min}\leq T \leq T_{max}$,
and thus the integration range has to be such that $T$ lies in this
interval.
The total number of micro-lensing events is then given by
\begin{equation}
N_{ev}=\int dN_{ev}~\epsilon(T)~
,\label{eq:th}
\end{equation}
where the integration is over the full range of
$d\mu du dv_T dx$. $\epsilon(T)$ is determined 
experimentally \cite{kn:Pratt,kn:MACHO}.
$T$ is related in a complicated way
to the integration variables,
because of this, no direct
analytical integration in eq.(\ref{eq:th}) can be performed.

To evaluate eq.(\ref{eq:th}) we define
an efficiency function $\epsilon_0(\mu)$
\begin{equation}
\epsilon_0(\mu) \equiv \frac{\int d N^{\star}_{ev}(\bar\mu)~ 
\epsilon(T)}
{\int d N^{\star}_{ev}(\bar\mu)}~,
\end{equation}
which measures the fraction of the total number of microlensing events
that meet the condition on $T$ at a
fixed MACHO mass $M=\bar\mu M_{\odot}$.
We now can write the total number of events in
eq.(\ref{eq:th}) as
\begin{equation}
N_{ev}=\int dN_{ev}~\epsilon_0(\mu)~.\label{eq:tl}
\end{equation}
Due to the fact that
$\epsilon_0$ is a function of $\mu$ alone, the integration in
$d\mu du dv_T dx$ factorizes into four integrals with independent
integration limits. 

The average lensing duration can be defined as follows
\begin{equation}
< T > = \frac{1}{\Gamma}~\int d\Gamma~T(x,\mu,v_T)~,
\end{equation}
where $T(x,\mu,v_T) = R_E(x,\mu)/v_T$. One easily finds that $< T >$
satisfies the following relation
\begin{equation}
< T > = \frac{2 \tau_{opt}}{\pi \Gamma} ~u_{TH}~.
\end{equation}

In order to quantify the expected number of events it is convenient
to take as an example a delta function distribution for the mass.
The rate of microlensing
events with
$A \geq A_{min}$ (or $u \leq u_{max}$), is then
\begin{equation}
\Gamma(A_{min})=u_{max} \tilde\Gamma = u_{max} 
D r_E \sqrt{\pi}~v_H \frac{\rho_0}{M_{\odot}}\frac{1}{\sqrt{\bar \mu}}
\int^1_0 dx[x(1-x)]^{1/2} H(x)~.\label{eq:ta}
\end{equation}

Inserting the numerical values for the LMC
(D=50~kpc and $\alpha=82^0$) we get
\begin{equation}
\tilde\Gamma=4
\times 10^{-13}~\left( \frac{v_H}{210~km/s}\right)
  \left( \frac{\rho_0}{0.3~GeV/cm^3}\right)
\frac{1}{\sqrt{M/M_{\odot}}}\ ~{\rm s^{-1}}.
\label{eq:tb}
\end{equation}
For an experiment monitoring $N_{\star}$ stars during an
observation time $t_{obs}$ the total number of events with a
magnification $A \geq A_{min}$ is:
$N_{ev}(A_{min})=N_{\star} t_{obs} \Gamma(A_{min})$.
In Table 1 we show some values of $N_{ev}$ for the LMC,
taking
$t_{obs}=1$ year, $N_{\star}=10^6$ stars and 
$A_{min} = 1.34$ (or $\Delta m_{min} = 0.32$).
\vskip 0.3 cm 
\begin{table*}[t!]
\caption{ }
\label{table1}
\begin{tabular}{cccc}
MACHO mass in $M_{\odot}$ & Mean $R_E$ in km & Mean microlensing time &
$N_{ev}$ \\
\hline
$10^{-1}$ & $0.3\times 10^9$ & 1 month & 4.5  \\
$10^{-2}$ & $10^8$ & 9 days & 15 \\
$10^{-4}$ & $10^7$ & 1 day & 165 \\
$10^{-6}$ & $10^6$ & 2 hours & 1662 \\
\hline
\end{tabular}
\end{table*}

\vskip 0.2 cm

\subsection*{Mass moment method}

A more systematic way to extract information on the masses is to use the
method of mass moments \cite{kn:Derujula,kn:Jetzer1,kn:Jetzer2}. 
The mass moments $<\mu^m>$ are defined as
\begin{equation}
<\mu^m>=\int d\mu~ \epsilon_n(\mu)~ 
\frac{dn_0}{d\mu}\mu^m~. \label{eqno:10}
\end{equation}
$<\mu^m>$ is related to $<\tau^n>=\sum_{events} \tau^n$,
with $\tau \equiv (v_H/r_E) T$, as constructed
from the observations and which can also be computed as follows
\begin{equation}
<\tau^n>=\int dN_{ev}~ \epsilon_n(\mu)~
\tau^n=V u_{TH} ~\gamma(m) <\mu^m>~,
\label{eqno:111}
\end{equation}
with $m \equiv (n+1)/2$.
For targets in the LMC $\gamma(m) = \Gamma(2-m) \widehat H(m)$ and
\begin{equation}
V \equiv 2 N_{\star} t_{obs}~ D~ r_E~ v_H=2.4 \times 10^3~ pc^3~ 
\frac{N_{\star} ~t_{obs}}{10^6~ {\rm star-years} }~, \label{eqno:121}
\end{equation}
\begin{equation}
\Gamma(2-m) \equiv \int_0^{\infty} \left(\frac{v_T}{v_H}\right)^{1-n}
f(v_T) dv_T~,
\label{eqno:131}
\end{equation}
\begin{equation}
\widehat H(m) \equiv \int_0^1 (x(1-x))^m H(x) dx~.  \label{eqno:14}
\end{equation}
The efficiency $\epsilon_n(\mu)$ is determined as follows 
\cite{kn:Derujula}
\begin{equation}
\epsilon_n(\mu) \equiv \frac{\int d N^{\star}_{ev}(\bar\mu)~ 
\epsilon(T)~ \tau^n}
{\int d N^{\star}_{ev}(\bar\mu)~ \tau^n}~, \label{eqno:15}
\end{equation}
where $d N^{\star}_{ev}(\bar\mu)$ is defined as $d N_{ev}$ 
in eq.(\ref{eq:th}) with
the MACHO mass distribution concentrated at a fixed mass
$\bar\mu$: $dn_0/d\mu=n_0~ \delta(\mu-\bar\mu)/\mu$. 
$\epsilon(T)$ is the experimental detection efficiency.
For a more detailed discussion on the efficiency see ref. \cite{kn:Masso}.

A mass moment $< \mu^m >$ is thus related to 
$< \tau^n >$ as given from the measured values 
of $T$ in a microlensing experiment by
\begin{equation}
< \mu^m > = \frac{< \tau^n >}{V u_{TH} \gamma(m)}~.
\label{eqno:16}
\end{equation}
The mean local density of MACHOs (number per cubic parsec)
is $<\mu^0>$. The average local mass density in MACHOs is
$<\mu^1>$ solar masses per cubic parsec.
 
The  mean mass, which we get from
the six events
detected by the MACHO team during their first two years, 
is \cite{kn:je} 
\begin{equation}
\frac{<\mu^1>}{<\mu^0>}=0.27~M_{\odot}~.
\label{eqno:aa}
\end{equation}
When taking for the duration $T$ the values corrected for ``blending'',
we get as average mass 0.34 $M_{\odot}$.
If we include also the two EROS events we get a value
of 0.26 $M_{\odot}$ for the mean mass (without taking into account
blending effects).
The resulting mass depends on the parameters
used to describe the standard halo model. In order to check this
dependence we varied the parameters of the standard halo model within
their allowed range and found
that the average mass changes at most by $\pm$ 30\%, which shows
that the result is rather robust. 
Although the value for the average mass we find with the mass moment
method is marginally consistent with the result of the MACHO team,
it definitely favours a lower average MACHO mass.

One can
also consider other models with more general
luminous and dark matter distributions, e.g. ones with a flattened halo
or with anisotropy in velocity space \cite{kn:Ingrosso},
in which case the resulting
value for the average mass would decrease significantly.

Another important quantity to be determined is the fraction $f$ of the local
dark mass density (the latter one given by $\rho_0$) detected
in the form of MACHOs, which is given by
$f \equiv {M_{\odot}}/{\rho_0} \sim 126~{\rm pc}^3$ $<\mu^1>$.
Using the values given by the MACHO collaboration
for their two years data \cite{kn:Pratt} 
we find $f \sim 0.54$,
again by assuming a standard spherical halo model.

Once several moments $< \mu^m >$ are known one can
get information on the mass distribution $dn_0/d\mu$. 
Since at present only few events towards the LMC are at disposal the 
different moments (especially the higher ones) can 
be determined only approximately.
Nevertheless, the results obtained so far
are already of interest and it is clear that in a few years,
due also to the new experiments under way (such as EROS II, OGLE II
and MOA in addition to MACHO),
it will be possible to draw more firm conclusions.

\section*{Present status of microlensing research}

It has been pointed out by Paczy\'nski \cite{kn:Paczynski} that microlensing 
allows the detection of MACHOs located in the galactic halo in the mass
range \cite{kn:Derujula1}  
$10^{-7} < M/M_{\odot} <  1$, as well as MACHOs in the disk
or bulge of our Galaxy \cite{kn:Paczynski1991,kn:Griest2}.
Since this first proposal microlensing searches have turned very quickly
into reality and in about a decade they have become an important
tool for astrophysical investigations. 
Microlensing is also very promising for the search of
planets around other stars in our Galaxy and generates  
large databases for variable stars, a field which has already
benefitted a lot. Because of the
increase of observations, since several
new experiments are becoming operative, the situation is changing
rapidly and, therefore,
the present results should be considered as preliminary.

\subsection*{Towards the LMC and the SMC}

In September 1993 the French collaboration EROS \cite{kn:Aubourg}
announced the discovery of 2 microlensing candidates
and the American--Australian
collaboration MACHO of one candidate \cite{kn:Alcock}
by monitoring stars in the LMC.

In the meantime the MACHO team reported the observation of
altogether 8 events
(one is a binary lensing event) analysing
their first two years of data
by monitoring about 8.5 million 
of stars in the LMC \cite{kn:Pratt}. 
The inferred optical depth is $\tau_{opt} = 2.1^{+1.1}_{-0.7} \times 10^{-7}$
when considering 6 events \footnote{In fact, the two disregarded events are 
a binary lensing and one which is rated as marginal.} 
(or $\tau_{opt} = 2.9^{+1.4}_{-0.9} \times 10^{-7}$ when
considering all the 8 detected events). Correspondingly, this implies
that about 45\% (50\% respectively) of the halo dark matter is in form of 
MACHOs and they find an average mass $0.5^{+0.3}_{-0.2} M_{\odot}$ 
assuming a standard spherical halo model.
It may well be that there is also a contribution of events due
to MACHOs located in the LMC itself or in a thick disk of our galaxy,
in which case the above results will change quite substantially.
In particular for the binary event there is evidence that
the lens is located in the LMC.
It has been estimated that the 
optical depth for lensing due to MACHOs in the LMC
or in a thick disk is about
$\tau_{opt}=5.4 \times 10^{-8}$ \cite{kn:Pratt}. However, this value
is model dependent so that at present it is not clear which fraction of the
events are due to self-lensing in the LMC (and similarly for the SMC).

Other events have been detected towards the LMC by the MACHO group, which 
have been put on their list of alert events. The full analysis of
the 1996 - 1998 seasons is still not published.

EROS has also searched for very-low mass MACHOs by looking for
microlensing events with time scales ranging from 30 minutes to 
7 days \cite{kn:EROS}. The lack of candidates in this range 
places significant constraints on any model for the halo that relies
on objects in the range $5 \times 10^{-8} < M/M_{\odot} < 2 \times 10^{-2}$.
Indeed, such objects may make up at most 20\% of the halo dark matter
(in the range between $5 \times 10^{-7} < M/M_{\odot} < 2 \times
10^{-3}$ at most 10\%). 
Similar conclusions have also been reached by the MACHO group
\cite{kn:Pratt}. 

Recently, the MACHO team reported \cite{kn:Alcock2} the first 
discovery of a 
microlensing event towards the Small Magellanic Cloud (SMC). The full
analysis of the four years 
data on the SMC is still underway, so that more candidates
may be found in the near future. A rough estimate 
of the optical depth leads to about the same value as found
towards the LMC.
The same event has also been observed by the EROS \cite{kn:eros}
and the Polish-American OGLE collaboration \cite{kn:ogle}. A second event
has been discovered in 1998 and found to be due to a binary
lens. This event has been followed by the
different collaborations, so that the combined data lead to a quite
accurate light curve, from which it is possible to get an upper limit for the
value of the proper motion of the lens \cite{kn:45,kn:46}. 
The result indicate that the lens system
is most probably located in the SMC itself, in which case the lens may be an 
ordinary binary star. It is remarkable that both the binary events
detected so far are due
to lenses in the Clouds themselves, making it plausible that this is the
case for the other lenses as well. 

Since the middle of 1996 the EROS group has put into operation a new
1 meter telescope, located in La Silla (Chile), and which is fully
dedicated to microlensing searches using CCD cameras. The improved experiment
is called EROS II.

\subsection*{Towards the galactic centre}

Towards the galactic bulge the 
Polish-American team OGLE \cite{kn:Udalski}
announced his first event also in September 1993.
Since then OGLE found in their data from the 1992 - 1995 observing
seasons altogether 18 microlensing events (one being a binary lens).
Based on their first 9 events the OGLE team estimated the optical depth
towards the bulge as \cite{kn:udal}  $\tau_{opt} = 
(3.3 \pm 1.2) \times 10^{-6}$.
This has to be compared with the theoretical calculations which
lead to a value \cite{kn:Paczynski1991,kn:Griest2} 
$\tau_{opt} \simeq (1 - 1.5)\times 10^{-6}$, which does, however,
not take into account the contribution of lenses in the bulge itself,
which might well explain the discrepancy. In fact, when taking into
account also the effect of microlensing by galactic bulge stars
the optical depth gets bigger \cite{kn:Kiraga}
and might easily be compatible with the measured value.
This implies the presence of a bar in the galactic centre.  
In the meantime the OGLE group got a new dedicated 1.3 meter
telescope located at the Las Campanas Observatory. The OGLE-2 collaboration
has started the observations in 1996 and is monitoring the
bulge, the LMC and the SMC as well.

The French DUO \cite{kn:Alard} team found 12 microlensing events (one of
which being a binary event) by monitoring the galactic bulge during the 1994
season with the ESO 1 meter Schmidt telescope. The photographic plates were
taken in two different colors to test achromaticity.
The MACHO  
\cite{kn:MACHO} collaboration 
found by now more than $\sim$ 150 
microlensing events towards the galactic bulge, most of
which are listed among the alert events, which are 
constantly updated \footnote{Current information on the MACHO
Collaboration's Alert events is maintained at the WWW site
http://darkstar.astro.washington.edu.}.
They found also
3 events by monitoring the spiral arms in the region of Gamma Scutum.
During their first season they found 45 events towards the bulge.
The MACHO team detected also in a long duration event the parallax
effect due to the motion of the Earth around the Sun \cite{kn:??}.
The MACHO first year data leads to an estimated optical depth of
$\tau_{opt} \simeq 2.43^{+0.54}_{-0.45} \times 10^{-6}$, which is roughly
in agreement with the OGLE result, and which also implies the presence
of a bar in the galactic centre.
These results are very important
in order to study the structure of our Galaxy. In this respect the measurement
towards the spiral arms will give important new information.
 
Some globular clusters lie in the galactic disk about half-way between us
and the galactic bulge. If globular clusters contain MACHOs, the latter can also
act as lenses for more distant stars located in the bulge.
Recently, we have analysed the microlensing events towards the galactic bulge,
which lie close to three globular clusters and found evidence that some
microlensing events are indeed due to MACHOs located in the globular clusters
\cite{kn:wandeler}. If this finding is confirmed, once more data will be
available, it would imply that also globular clusters contain an important 
amount of dark matter in form of MACHOs, which probably would be brown dwarfs 
or white dwarfs.

\subsection*{Towards the Andromeda galaxy}

Microlensing searches have also been conducted towards M31,
which is an interesting
target \cite{kn:Crotts,kn:Baillon,kn:Jetzer}. In this
case, however, one has to use the so-called ``pixel-lensing'' method,
since the source stars are in general no longer resolvable. Two groups
have performed searches: the French AGAPE \cite{kn:Agape}
using the 2 meter telescope at Pic du Midi  
and the American VATT/COLUMBIA \cite{kn:VATT}, 
which used the 1.8 meter VATT-telescope
located on Mt. Graham and the 4 meter KNPO telescope.
Both teams showed that the pixel-lensing method works, however,
the small amount of observations done so far does not allow
to draw firm conclusions. The VATT/COLUMBIA
team found six candidates
which are consistent with microlending, however, additional observations
are needed to confirm this.
Pixel-lensing could also lead to the discovery of microlensing
events towards the M87 galaxy, in which case the best would be to use
the Hubble Space Telescope \cite{kn:M87}. It might also
be interesting to look towards dwarf galaxies of the local group.

\subsection*{Further developments}

A new collaboration between New Zealand and Japan, called MOA, started
in june 1996 to perform observations using the 0.6 meter telescope
of the Mt. John Observatory \cite{kn:Moa}. 
The targets are the LMC and the galactic
bulge. They will in particular search for short timescale ($\sim$
1 hour) events, and will then be particularly sensitive to objects
with a mass typical for brown dwarfs.

It has to mentioned that there are also collaborations between 
different observatories (for instance PLANET \cite{kn:PLANET}
and GMAN \cite{kn:GMAN}) 
with the aim to perform
accurate photometry on alert microlensing events. 
The GMAN collaboration was able to accurately get photometric data
on a 1995 event towards the galactic bulge. The light curve
shows clearly a deviation due to the extension of the source star
\cite{kn:gman}.
A major goal of the PLANET and GMAN collaborations 
is to find planets in binary microlensing events
\cite{kn:Mao,kn:Loeb,kn:Rhie}.
Moreover, microlensing searches are also very powerful ways to 
get large database for the study and discovery of many variable stars.

At present the only information available from a microlensing event
is the time scale, which depends on three parameters: distance,
transverse velocity and mass of the MACHO. A possible way to get more
information is to observe an event from different locations, with typically
an Astronomical Unit in separation. This could be achieved 
by putting a parallax satellite into solar orbit \cite{kn:Refsdal,kn:Gould}.

The above list of presently active collaborations and main results
shows clearly that this field is just at the beginning and that
many interesting results will come in the near future.

\section*{Formation of dark clusters}
We turn now to the discussion of a scenario for the formation
of dark clusters of MACHOs and cold molecular clouds.
Our scenario \cite{kn:de1,kn:de2,kn:de4}
encompasses the one originally proposed by Fall and Rees \cite{kn:fall} to 
explain the origin of globular clusters and can be summarized as follows.
After its initial collapse, the proto galaxy (PG) is expected to be shock
heated to its virial temperature $\sim 10^6$ K. Because of thermal 
instability, density enhancements rapidly grow as the gas cools. 
Actually, overdense regions cool
more rapidly than average, and so proto globular cluster
(PGC) clouds form in pressure equilibrium with hot diffuse gas. When the 
PGC cloud temperature reaches $\sim 10^4$ K, hydrogen recombination occurs: 
at this stage, their mass and size are
$\sim 10^5 (R/{\rm kpc})^{1/2} M_{~\odot}$ and $\sim 10~(R/kpc)^{1/2}$ pc,
respectively ($R$ is the galactocentric distance). Below $10^4$ K, the 
main coolants are $H_2$ molecules and any 
heavy element produced in a first chaotic galactic phase. 
The subsequent evolution of the PGC clouds will be very different 
in the inner and outer part of the Galaxy, depending on the decreasing 
ultraviolet (UV) flux as the galactocentric distance $R$ increases.

As is well known, in the central region of the Galaxy
an Active Galactic Nucleus (AGN) and a first population of  
massive stars are expected to form, which act as strong sources of UV 
radiation that dissociates the $H_2$ molecules. It is not difficult to 
estimate that $H_2$ depletion should happen for galactocentric distances 
smaller than $10-20$ kpc. As a consequence,
cooling is heavily suppressed in the inner halo, and so the PGC clouds here
remain for a long time at temperature $\sim 10^4$ K, resulting in the imprinting
of a characteristic mass $\sim 10^6 M_{\odot}$. Eventually, the UV flux will
decrease, thereby permitting the formation of $H_2$.
As a result, the cloud temperature drops below $\sim 10^4$ K and the subsequent 
evolution leads to star formation and ultimately to globular clusters.
 
Our main point is that in the outer halo -- namely for galactocentric 
distances larger than $10-20$ kpc -- no substantial $H_2$ depletion should 
take place (owing to the distance suppression of the UV flux).
Therefore, the PGC clouds cool and contract. When their number density 
exceeds $\sim 10^8$ cm$^{-3}$, further $H_2$ is produced via 
three-body reactions
($H+H+H \rightarrow H_2+H$ and $H+H+H_2 \rightarrow   2H_2$), which makes 
in turn the cooling efficiency increase dramatically.
This fact has three distinct implications:
(i) no imprinting of a characteristic PGC cloud mass shows up,
(ii) the Jeans mass can drop to values considerably smaller
than $\sim 1~M_{\odot}$, and
(iii)  the cooling time is much shorter than the free-fall time.
In such a situation a 
subsequent fragmentation occurs into smaller and smaller clouds that remain 
optically thin to their own radiation. The process stops when the clouds 
become optically thick to their own line emission -- this happens when the
Jeans mass is as low as $\sim 10^{-2}~M_{\odot}$. In this manner, dark
clusters should form, which contain brown dwarfs in the mass range 
$10^{-2}-10^{-1}~M_{\odot}$.

Before proceeding further, two observations are in order. First, it seems 
quite natural to suppose that -- much in the same way as it occurs for 
ordinary stars -- also in this case the fragmentation process that gives 
rise to individual brown dwarfs should produce a substantial fraction of 
binary brown dwarfs. It is important 
to keep in mind that the mass fraction of primordial 
binaries can be as large as $50\%$. Hence, we see that MACHOs consist of 
both individual and binary brown dwarfs in the present scenario
\cite{kn:mnras,kn:apj}. Second,
we do not expect the fragmentation process to be able to convert the whole 
gas in a PGC cloud into brown dwarfs. For instance, standard stellar formation 
mechanisms lead to an upper limit  of at most $40\%$ for 
the conversion efficiency. Thus, a substantial fraction $\tilde f$ 
of the primordial 
gas -- which is mostly $H_2$ -- should be left over. Because brown dwarfs 
do not give rise to stellar winds,
this gas should remain confined within a dark 
cluster. So, also cold $H_2$ self-gravitating clouds should presumably be 
clumped into dark clusters, along with some residual diffuse gas (the 
amount of diffuse gas inside a dark cluster has to be low, for otherwise it 
would have been observed in optical and radio bands).

Unfortunately, the total lack of any observational information about
dark clusters would make any effort to understand their structure and
dynamics practically impossible, were it not for some remarkable
insights that our unified treatment of globular and dark clusters provides 
us. In the first place, it looks quite natural to assume that also dark
clusters have a denser core surrounded by an extended spherical
halo. Moreover, in the lack of any further information it seems reasonable
to suppose (at least tentatively) that the dark clusters have the same
average mass density as globular clusters. Hence, we obtain
$r_{DC}\simeq 0.12~ ({M_{DC}}/{M_{\odot}})^{1/3}$  pc,
where $M_{DC}$ and $r_{DC}$ denote the mass and the median radius of a
dark cluster, respectively.
As a further implication of the above scenario, we stress that -- at
variance with the case of globular clusters -- the initial mass
function of the dark clusters should be smooth, since the monotonic
decrease of the PGC cloud temperature fails to single out any
particular mass scale. In addition, the absence of a quasi-hydrostatic
equilibrium phase for the dark clusters naturally suggests 
$M_{DC}\leq 10^6~M_{\odot}$. Finally, we suppose for definiteness 
that all brown dwarfs have 
mass $\simeq 0.1~M_{\odot}$, while the molecular cloud spectrum will be 
taken to be $10^{-3}~M_{\odot}\leq M_m \leq 10^{-1}~M_{\odot}$.

\section*{Observational tests}
We list schematically some observational tests for the present scenario.

{\bf Clustering of microlensing events} -- The most promising way to detect 
dark clusters is via correlation effects in microlensing observations, as 
they are expected to exhibit a cluster-like distribution \cite{kn:maoz}. 
Indeed, it has 
been shown that a relatively small number of microlensing events would be 
sufficient to rule out this possibility, while to confirm it more events 
are needed. However, we  have seen that core collapse can liberate a 
considerable fraction of MACHOs from the less massive clusters, and so an 
unclustered MACHO population is expected to coexist with dark clusters in 
the outer halo -- detection of unclustered MACHOs would therefore not 
disprove the present model.

{\bf $\gamma$-rays from halo clouds} -- A signature for the presence of 
molecular clouds in the galactic halo should be a $\gamma$-ray flux 
produced in the scattering of high-energy cosmic-ray protons on $H_2$
\cite{kn:de1,kn:de2}.
As a matter of fact, an essential
ingredient is the knowledge of the cosmic ray flux in the halo. Unfortunately,
this quantity is unknown and the only available 
information comes from theoretical estimates.
Moreover, we
assume the same energy distribution of the cosmic rays as measured on Earth.
The presence of
magnetic fields in the halo is expected to give rise to a temporary 
confinement of 
cosmic ray protons similar to what happens in the disk. In addition, there
can also be sources of cosmic ray protons located in the halo itself, as for
instance isolated or binary pulsars in globular clusters.
The best chance to detect the $\gamma$-rays in question is provided
by observations at high galactic latitude.
We find that - regardless of the adopted value for the flatness of
the halo -
at high-galactic 
latitude $\Phi_{\gamma}^{~\rm DM}(>1{\rm ~GeV})$ lies in the range
$\simeq 6-8 \times 10^{-7}$ $\gamma$ cm$^{-2}$ s$^{-1}$ sr$^{-1}$
(assuming a fraction $\tilde f \simeq 0.5$ for the dark matter in form of 
cold clouds). 
However, the shape of the contour lines strongly depends on
the flatness parameter \cite{kn:gamma}. 

A few months ago, Dixon et al. \cite{dixon} have re-analyzed the EGRET data 
concerning the diffuse $\gamma$-ray flux with a wavelet-based technique. 
After subtraction of the isotropic extragalactic component and of the 
expected contribution from the Milky Way, they find a statistically 
significant diffuse emission from the galactic halo. At high-galactic 
latitude, the integrated halo flux above 1 GeV turns out to be 
$\simeq 10^{-7}-10^{-6}$ $\gamma$ cm$^{-2}$ s$^{-1}$ sr$^{-1}$, 
which is slightly less than the 
diffuse extragalactic flux (Sreekumar et al. \cite{sreekumar}).
Our estimate for the halo $\gamma$-ray flux turns out to be
in remarkably good agreement 
with the discovery by Dixon et al. \cite{dixon}.
The next generation of $\gamma$-ray satellites like AGILE and
GLAST will be able  to test our model, 
thanks to their better angular resolution.

{\bf CBR anisotropy} -- An alternative way to discover the molecular 
clouds under consideration relies upon their emission in the microwave band
\cite{di}. 
The temperature of the clouds has to be close to that of the cosmic background
radiation (CBR). Indeed, an upper limit of $\Delta T/T \sim 10^{-3}$ can
be derived by considering the anisotropy they would
introduce in the CBR due to their higher temperature. Realistically, molecular
clouds cannot be regarded as black body emitters because they mainly 
produce a set of molecular rotational transition lines. If we consider
clouds with cosmological primordial composition, then the only molecule that
contributes to the microwave band with optically thick 
lines is $LiH$, whose lowest rotational
transition occurs at $\nu_0 = 444$ GHz with broadening $\sim 10^{-5}$
(due to the turbulent velocity of molecular clouds in dark clusters). 
This line would be detectable using the Doppler shift effect.
To this aim, it is convenient to consider M31 galaxy, for
whose halo we assume the same picture as outlined above for our 
galaxy. Then we expect that molecular 
clouds should have typical rotational speeds of 50-100 km s$^{-1}$. 
Given the fact that the clouds possess a peculiar velocity
(with respect to the CBR) the 
emitted radiation will be Doppler shifted, with 
$\Delta\nu /\nu_{0}\sim\pm ~10^{-3}$.
However, the precise chemical composition of molecular clouds in the
galactic halo is unknown. Even if the heavy
molecule abundance is very low (as compared with the abundance in
interstellar clouds), many optically thick lines corresponding to the lowest
rotational transitions would show up in the microwave band. In this case, 
it is more convenient to perform broad-band measurements and the Doppler shift 
effect results in an anisotropy in the CBR. 
Since it is difficult to work with fields of view
of a few arcsec, we propose to measure the 
CBR anisotropy between two fields of view - on opposite
sides of M31 - separated by $\sim 4^0$ and with angular resolution
of $\sim 1^0$. We suppose that the halo of M31 consists of
$\sim 10^6$ dark clusters which lie within 25-35 kpc.
Scanning an annulus of $1^0$ width and internal angular
diameter $4^0$, centered at M31, in 180 steps of $1^0$, we would find
anisotropies of $\sim 10^{-5} ~\tilde f~ \bar\tau$ in $\Delta T/T$. Here, most of 
the uncertainties arise from the estimate of the
average optical depth $\bar\tau$, which mainly depends
on the molecular cloud composition. In conclusion, since the theory does not
allow to establish whether the expected anisotropy lies above or below
current detectability ($\sim 10^{-6}$), only observations can resolve this
issue.

{\bf Absorption-lines} -- Cold clouds clumped into dark clusters 
can also be observed through absorption lines (due to heavy 
molecules) both in UV and in optical bands in the spectra of LMC stars, 
which lie very close (within $ \sim 1'$) to a previously microlensed one. 
 
{\bf Infrared searches} -- Another possibility of detecting MACHOs
is via their infrared emission \cite{di}. 
In order to be specific, let us assume that all
MACHOs have same mass 0.08 $M_{\odot}$ and age
$10^{10}$ yr. Accordingly, their surface temperature is
$\sim 1.4 \times 10^3$ K and they emit most of their radiation
(as a black body) at $\nu_{max} \sim 11.5 \times 10^{13}$ Hz.
First, we consider MACHOs  located in M31.
In this case, we find a surface brightness 
$I_{\nu_{max}} \sim 2.1 \times 10^3~(1-\tilde f) $ Jy sr$^{-1}$ and
$0.5 \times 10^{3}~(1-\tilde f)$ Jy sr$^{-1}$
for projected separations from the M31 center $b=20$ kpc and 40 kpc, 
respectively. Although these values are about one order of magnitude below the 
sensitivity of the detectors on ISO Satellite, they lie above the threshold of 
the detector abord the future planned SIRFT Satellite.
For comparison, we recall that the halo of our galaxy would have in
the direction of the galactic pole a surface brightness
$I_{\nu_{max}}\sim 2 \times 10^{3}~{\rm Jy~sr^{-1}}$, provided 
MACHOs make up the total halo dark matter.
Nevertheless, the infrared radiation
originating from  MACHOs in the halo of our galaxy can be recognized (and
subtracted) by its characteristic angular modulation.
Also, the signal from the
M31 halo can be identified and separated from the galactic background via its
b-modulation. 
Next, we point out that the angular size of dark clusters in the
halo of our galaxy at a distance of $\sim 20$ kpc is $\sim 1.8'$ and the
typical separation among them is
$\sim 14'$. As a result, a characteristic pattern of
bright (with intensity $\sim 3\times 10^{-2}$ Jy at $\nu_{max}$ within
angular size $1.8'$) and dark spots 
should be seen by pointing the detector into
different directions.

\section*{Conclusions}

The mistery of the dark matter is still unsolved, however, thanks
to the ongoing microlensing and pixel-lensing
experiments there is hope that
progress on its nature in the galactic halo
can be achieved within the next few year.
An important point will be to determine whether the MACHOs are in the halo
or rather in the LMC or SMC themselves, as suggested by the binary lens events.

Substantial progress will also be done in the study of the
structure of our Galaxy and this especially once data from the
observations towards the spiral arms will be available.
Microlensing is also very promising for the discovery of planets.
Although being a rather young observational technique
microlensing has
already allowed to make substantial progress and
the prospects for further contribution
to solve important astrophysical problems look very bright. 

It has also to be mentioned that
only a fraction of the halo
dark matter might be in form of MACHOs,
in which case there is the problem of explaining the nature
of the remaining dark matter and the formation of the MACHOs.
Before invoking the need for new particles
as galactic dark matter candidates for the remaining fraction, 
one should seriously consider the
possibility that it is in the form of cold molecular clouds.
Several observational methods
have been proposed to test this scenario,
in particular via the induced $\gamma$-ray flux for which the predicted
value is in remarkably good agreement with the measurement
of EGRET \cite{dixon}. \\

\noindent
{\bf Acknowledgements}\\

I would like to thank B. Paczy\'nski for an important comment 
and for bringing to my attention several recent papers.

\end{document}